\documentstyle[12pt]{article}
\newcommand{\del}{\frac{\Delta} 2 }
\newcommand{\tg}{${\cal T}$ }

\begin{document}
\title{
Elliptic Ruijsenaars-Schneider model from 
the cotangent bundle
over the two-dimensional current group.
}
\author{G.E.Arutyunov,
\thanks{Steklov Mathematical Institute,
Vavilov 42, GSP-1, 117966, Moscow, Russia; arut@class.mi.ras.ru}\\
S.A.Frolov
\thanks{Steklov Mathematical Institute,
Vavilov 42, GSP-1, 117966, Moscow, Russia; frolov@class.mi.ras.ru}\\
and\\
P.B.Medvedev \thanks
{Institute of Theoretical and Experimental Physics,
 B.Cheremushkinskaja 25, 117259 Moscow, Russia}
}
\date {}
\maketitle
\begin{abstract}
It is shown that the elliptic Ruijsenaars-Schneider model can be obtained 
from the cotangent bundle
over the two-dimensional current group
by means of the Hamiltonian reduction 
procedure. 
\end{abstract} 
\newpage

\section{Introduction}
In our recent paper \cite{afm} we have shown that the 
elliptic Ruijsenaars-Schneider (RS) model \cite{R} can be obtained 
by means of the Poisson reduction technique from the affine 
Heisenberg double.  The aim of the present note is to derive the same 
RS model from the cotangent bundle over the two-dimensional centrally 
extended current group $GL(N)(z,\bar z)$ applying this time the 
Hamiltonian reduction procedure \cite{Arn}-\cite{AT}. It is 
worthwhile to note that the cotangent bundle over the centrally 
extended  $sl$ current algebra was used in \cite{GNH,AM} to obtain 
the elliptic Calogero-Moser model.  In this short note we shall not 
discuss the state of affairs in the problem, see \cite{afm} and 
references therein. 

The plan of the paper is as follows. In the second section we briefly  
describe an infinite-dimensional phase space, which can be regarded 
as the cotangent bundle 
over the two-dimensional centrally extended current group.  
Then we fix the momentum map, corresponding to the 
natural action of the group and characterize the reduced phase space. 
The resulting $L$-operator appears to be equivalent to the 
$L$-operator of the RS model. 
In the third section 
we calculate the Poisson bracket of the reduced phase 
space variables
and prove that it coincides with the one of the 
RS model. The reduction procedure leads to the dynamical $r$-matrix 
which just as in our previous paper is equivalent to the one 
obtained in \cite{Sur1}.

\section {Cotangent bundle over $GL(N)(z,\bar z)$}
The cotangent bundle ${\cal T}$ over the two-dimensional centrally 
extended current group $GL(N)(z,\bar z)$ is a straightforward 
generalization of the cotangent bundle over affine $\widehat{GL(N)}$. The 
Poisson structure on \tg  is defined as follows. Let $A(x,y)=\sum 
A_{mn}e^{imx+iny}$ and $g(x,y)=\sum g_{mn}e^{imx+iny}$ be formal Fourier 
series in variables $x$ and $y$ with values in $gl(N)$ and $GL(N)$ 
respectively.  It is convenient to use variables $z=x+\frac\Delta{2\pi} y$ 
and $\bar{z}=x+\frac{\overline{\Delta}}{2\pi} y$, where $\Delta$ is a modular 
parameter with ${\mbox {Im}} \Delta >0$. In what follows we shall often use 
the notation $A(x,y)\equiv A(z)$ and $g(x,y)\equiv g(z)$.
The matrix elements $A_{mn}$ and $g_{mn}$ 
can be regarded as generators of the algebra of 
functions on \tg . In close analogy with $T^*\widehat{GL(N)}$ the Poisson 
structure can be written as 
\begin{eqnarray} \{ A_1(z),A_2(w)\} &=&\frac 1 
2 [P,A_1(z) -A_2(w)] \delta(z-w) -k P\frac{\partial}{\partial \bar z 
}\delta(z-w) \nonumber\\ 
 \{g_1(z),g_2(w)\} &=& 0\nonumber\\ 
\{ A_1(z),g_2(w)\}
&=&g_2 (w)P\delta (z-w)
\label{pb}
\end{eqnarray} 
where $k$ is a (fixed) central charge and 
$\delta(z)$ is the two-dimensional 
$\delta$-function.  Here we use a standard tensor notation and $P$ is 
the permutation operator.  

The action of $GL(N)(z,\bar z)$ on \tg 
\begin{eqnarray*} 
A(z)&\to &T^{-1}(z)A(z)T(z) +kT^{-1}(z)\bar\partial T(z),\\
g(z)&\to &T^{-1}(z)g(z)T(z)
\end{eqnarray*} 
is Hamiltonian. Thereby, we can consider the Hamiltonian reduction 
of \tg over the action of $GL(N)(z,\bar z)$.

The momentum map taking value in 
$gl(N)(z,\bar z)^*$ looks  as follows:
$$
M(z)=k\bar\partial g(z)g^{-1}(z)+A(z)-g(z)A(z)g^{-1}(z). 
$$ 
It is easy to check that $M(z)$ does generate 
the action of the current group.  We fix the value of $M(z)$ as:  
\begin{equation} 
M(z)=-\frac {k}{2{\mbox {Im}} \Delta}h+2\pi ik \delta_\varepsilon 
(z)\frac{1-e^{-ix}}{i} K.  
\label{m}
\end{equation} 
Here $h$ and 
$\varepsilon$ are arbitrary complex numbers, 
$$ 
\delta_\varepsilon (z)=
\frac {2\pi}{{\mbox {Im}} \Delta} \delta_\varepsilon (x) \delta (y), 
$$ 
$$ 
\delta_\varepsilon (x)=\frac 1 \varepsilon \left(
\theta (x+\frac\varepsilon 2) -\theta (x-\frac\varepsilon 2) \right)
=\frac1{2\pi i \varepsilon }\sum^{n=+\infty}_{n=-\infty} \frac1n 
(e^{in\frac\varepsilon 2} -e^{-in\frac\varepsilon 2}) e^{inx}, 
$$ 
and 
$K$ is a constant matrix $K=e\otimes e^t$, where $e$ is the 
$N$-dimensional vector with entries $e_i = 1/\sqrt{N}$.

To obtain a finite dimensional reduced phase space one has to 
consider the limit when $\varepsilon$ goes to zero. To treat this 
limit we employ the same strategy as in \cite{afm}.
We multiply the both sides of (\ref{m}) by
$ g(z)$, that gives
\begin{equation}
k\bar\partial g(z)+A(z)g(z)-g(z)A(z)
+\frac {k}{2{\mbox {Im}} \Delta}hg(z)=
2\pi ik \delta_\varepsilon (z)K\frac{1-e^{-ix}}{i}g(z) .  
\label{fe}
\end{equation}
The l.h.s. of this equation
does not have any explicit dependence on $\varepsilon$.
As to the r.h.s., when $\varepsilon$ tends to zero, 
$\delta_\varepsilon (z)$ becomes proportional to 
$\delta(x) \delta(y)$ and the 
r.h.s. is well defined only if the function 
$\frac{1-e^{-ix}}{i}g(x,0)$ is well 
defined at $x=0$. In this case 
$\lim_{\varepsilon \to 0} 
\delta_\varepsilon (z)\frac{1-e^{-ix}}{i}g(z)= 
\delta (z)Z$, where $Z=\frac{1-e^{-ix}}{i}g(x,0)|_{x=0}$.

So we define the constraint surface as being the solution of 
the equation
\begin{equation}
k\bar\partial g(z)+A(z)g(z)-g(z)A(z)+
\frac {k}{2{\mbox {Im}} \Delta}hg(z)=2\pi ik \delta (z)KZ .  
\label{fee}
\end{equation}
and in the following we shall explore solutions of this equation.

We start with the following differential equation
\begin{equation}
k\bar\partial g(z)+Dg(z)-g(z)D+\frac {k}{2{\mbox {Im}} \Delta}hg(z)=
2\pi ik \delta (z)Y,      
\label{e}
\end{equation}
where $D$ is a constant diagonal matrix and $Y$ is an arbitrary
constant matrix.

\noindent It is useful to introduce a function of two complex variables 
\begin{equation} 
W(z,s)= 
\frac{\sigma(z+s)}
{\sigma(z)\sigma(s)}
e^{-\frac{\zeta(\pi)}{\pi}zs}e^{is\frac{z-\bar z}{\Delta -\overline{\Delta}}}.
\nonumber
\end{equation} 
Here $\sigma (z)$ and $\zeta (z)$ are the Weierstrass $\sigma$- 
and $\zeta$-functions with periods equal to $2\pi$ and $\Delta$.
This function is the only doubly periodic solution of the following equation:
\begin{equation} 
\bar\partial 
W(z,s)+i\frac{s}{\Delta -{\overline\Delta}} W(z,s)=2\pi i\delta(z) \nonumber 
\end{equation} 

In terms of $W$ eq. (\ref{e}) can be solved as  
\begin{equation}
g(z)=\sum_{ij} \frac{\sigma(z+s_{ij})}
{\sigma(z)\sigma(s_{ij})}
e^{-\frac{\zeta(\pi)}{\pi}zs_{ij}}
e^{is_{ij}\frac{z-\bar z}{\Delta -\overline\Delta}}Y_{ij}E_{ij} =
\sum_{ij} W(z,s_{ij}) Y_{ij}E_{ij}.
\label{fin1}
\end{equation} 
Here we introduce the notation $s_{ij}=q_i -q_j +h$, $q_i=\frac{\Delta 
-\overline\Delta}{ik}D_i$ 

Now we turn to the momentum map equation (\ref{fee}).
By using a generic gauge 
transformation one can diagonalize the field $A$. Then equation (\ref{fee}) 
takes the form of eq.(\ref{e})
\begin{equation}
k\bar\partial g'(z)+Dg'(z)-g'(z)D
+\frac {k}{2{\mbox {Im}} \Delta}hg'(z)=2\pi ik \delta (z)K'Z',      
\nonumber
\end{equation}
where 
$$
A(z)=T(z)DT^{-1}(z)-k\bar\partial T(z)T^{-1}(z),~~
g(z)=T(z)g'(z)T^{-1}(z)
$$
for some $T$
and $Z'=xg'(x,0)|_{x=0}$. We also have
$$
K'=T^{-1}(0)KT(0)=T^{-1}(0)e\otimes e^tT(0)=f\otimes v^t,~~~<f,v>=1
$$
i.e. $f=T^{-1}(0)e$ and $e^tT(0)=v^t$.
According to (\ref{fin1}) we find
$$
g'(z)=\sum_{ij}
W(z,s_{ij})(K'Z')_{ij}E_{ij}.
$$
Taking the value of $xg'(x,0)$ at the point $x=0$ we arrive
at the compatibility condition
$$
Z'=K'Z'=f\otimes v^t Z',~~~<f,v>=1. 
$$
The solution of this equation is $Z'=f\otimes c^t$, where $c$ is an 
arbitrary vector.
Now it is easy to find $Z$:
$$Z=T(0)Z'T^{-1}(0)=T(0)f\otimes c^tT^{-1}(0)=
e\otimes c^tT^{-1}(0)
\equiv e\otimes b^t
$$
Thus, we get
\begin{equation}
k\bar\partial g(z)+A(z)g(z)-g(z)A(z)
+\frac {k}{2{\mbox {Im}} \Delta}hg(z)=2\pi ik(e\otimes e^t) (e\otimes b^t)\delta (z).
\label{fg11}
\end{equation}

To summarize, eq.(\ref{fee}) has a solution for any field $A$ 
and for any field $g$, such that $xg(x,0)|_{x=0}$ is of the form
$ e\otimes b^t$. For a fixed field $A$ and a vector $b$ this solution 
is unique.
Note that, in general,
$<b,e>\neq 1$. The form of the r.h.s. of (\ref{fg11}) shows that
the isotropy group of this equation is
$$
G_{isot}=\{T(z)\subset G(z,\bar z)~~|~~ T(0)e=\lambda e, 
\lambda\in {\bf C}\}.
$$ 
This group transforms a solution of 
(\ref{fg11}) into another one, so the reduced phase space is defined 
as 
$$
{\cal P}_{red}=\frac{\mbox{all solutions of (\ref{fee}})}{G_{isot}}.
$$
The group $G_{isot}$ is large enough to diagonalize the field
$A$ and hence we can parametrize the reduced phase space by the section
$(D,L)$, where $L$ is a solution of (\ref{fee}) with $A=D$. 
One can easily see that ${\cal P}_{red}$ is finite dimensional
and it's dimension is 
equal to $2N$, i.e. $N$ coordinates of $D$ plus $N$ coordinates of the 
vector $b$.
Due to eq.(\ref{fin1}) the corresponding $L$-operator has the following form:
\begin{equation}
L(z)=\sum_{ij}
\frac{\sigma(z+s_{ij})}
{\sigma(z)\sigma(s_{ij})}
e^{-\frac{\zeta(\pi)}{\pi}zs_{ij}}
e^{is_{ij}\frac{z-\bar z}{\Delta -\overline\Delta}}e_{i}b_j E_{ij} =
\sum_{ij} W(z,s_{ij}) Y_{ij}E_{ij}.
\label{l}\end{equation}
$L(z)$ and the $L$-operator obtained in \cite{afm} are related by 
the gauge transformation with the diagonal matrix
$e^{-i\frac{z-\bar z}{\Delta -\overline\Delta}q}$ and the shift of 
the spectral parameter: $z\to z-\del$.

The standard 
$L$-operator of the elliptic Ruijsenaars-Schneider model
can be obtained by
multiplying $L(z)$ by the function $
\frac {\sigma (z)\sigma(h)}{\sigma (z+h)}
e^{\frac{\zeta (\pi )}{\pi}zh-ih\frac{z-\bar z}{\Delta -\overline\Delta} }$
and performing the gauge transformation by means of the diagonal matrix $
e^{(\frac{\zeta (\pi )}{\pi}z-i\frac{z-\bar z}{\Delta -\overline\Delta})q} 
$:
\begin{equation}
L^{{\em Ruij}}(z)=
\frac {\sigma (z)\sigma(h)}{\sigma (z+h)}
e^{\frac{\zeta (\pi )}{\pi}zh-ih\frac{z-\bar z}{\Delta -\overline\Delta} } 
e^{(\frac{\zeta (\pi )}{\pi}z-i\frac{z-\bar z}{\Delta -\overline\Delta})q} 
L(z)
e^{-(\frac{\zeta (\pi )}{\pi}z-i\frac{z-\bar z}{\Delta 
-\overline\Delta})q} 
\label{rg}
\end{equation}

\section{The Poisson structure on the reduced space}

Our goal in this section is to examine the Poisson structure 
on the reduced phase space.  We should calculate the Poisson 
brackets for the coordinates $D$-s and $b$-s. Following the general 
Dirac procedure one should find a $G_{isot}$-invariant extension 
of functions on the reduced phase space ${\cal P}_{red}$ to a vicinity of 
${\cal P}_{red}$ and then calculate the Dirac bracket. 

The bracket for the coordinates $D_i$ and $b_i$ can 
be extracted from the bracket for $D$ and $L(z)$. 
The simplest gauge invariant extension for 
$D$ and $L(z)$ looks as follows:
\begin{equation}
D\rightarrow 
D[A]=T^{-1}(z)A(z)T(z)+kT^{-1}(z)\bar\partial T(z),
\label{fac}
\end{equation} 
\begin{equation}
L(z) \rightarrow {\cal L}[A,g](z)=T^{-1}(z)g(z)T(z)
\label{ll}
\end{equation} 
Some comments are in order. Eq.(\ref{fac}) is a solution of 
the factorization problem for $A(z)$. Generally this solution is not unique
but we fix the matrix $T[A]$ by the boundary condition 
$T[A](0)e= e$ that kills 
the ambiguity up to the action of the Weil group (see, e.g. \cite{AM}) and 
makes (\ref{fac}) to be correctly defined.  It is obvious that on ${\cal 
P}_{red}$: $T[A]=1$ and ${\cal L}[A,g](z)=L(z)$. 

The most interesting is the bracket for ${\cal L}(z)$ and 
${\cal L}(w)$ defined by eq.(\ref{ll}). At the end of the section
we shall comment on the  
contribution from the second class constraints to the Dirac bracket. 
By definition, one has 
\begin{eqnarray} 
\{{\cal 
L}_1,{\cal L}_2\}_{{\cal P}_{red}}&=& \left(\{T_1,T_2\}L_1L_2 -L_2 
\{T_1,T_2\}L_1-L_1 \{T_1,T_2\}L_2 \right. \nonumber \\ 
&+& L_1L_2 
\{T_1,T_2\} -\{T_1,g_2\}L_1-\{g_1,T_2\}L_2 \nonumber \\ 
&+&L_2\{g_1,T_2\}+L_1\{T_1,g_2\}
\left. \right) |_{{\cal P}_{red}}
\label{gh} 
\end{eqnarray}
Here we took into account that $T[A]|_{{\cal P}_{red}}=1$.

Let us first calculate
$$
\{g_{ij}(z),T_{kl}(w)\}=\sum_{m,n}
\int d^2 z' \{g_{ij}(z),A_{mn}(z')\} \frac{\delta T_{kl}(w)}
{\delta A_{mn}(z')}.  
$$ 
Performing the variation of 
the both sides of (\ref{fac}), we get
\begin{equation}
X(z)=t(z)D-Dt(z)-k \bar\partial t(z)+d,
\label{sd}
\end{equation}
where $X(z)=\delta A(z)$, $t(z)=\delta T(z)$
and $d=\delta D$.

\noindent The general solution of (\ref{sd}) is
\begin{equation}
t(z)=Q-\frac{1}{2\pi ik}\sum_{i, j}
\int d^2 w W(z-w,q_{ij})X_{ij}(w) E_{ij}.
\nonumber
\end{equation}
Here $Q$ is some constant diagonal matrix and we introduce the following 
notation 
$$ 
w(z,0)=\lim_{\varepsilon \to 
0}(w(z,\varepsilon) - \frac{1}{s}) 
=\zeta(z)-\frac{\zeta(\pi)}{\pi}z+i \frac{z-\bar{ z}}{\Delta-\overline{\Delta}}.
$$
This function solves the equation
$$
\bar{\partial}W(z,0)=2\pi i\delta(z) -\frac i{\Delta-\overline{\Delta}} 
$$
The solution $t(z)$ obeying the condition $t(0)e=0$ has the 
following form
\begin{equation}
t(z)=\frac1{2\pi ik}\sum_{i,j}\int d^2 w ( W(-w,q_{ij})
X_{ij}(w)E_{ii}-
W(z-w,q_{ij})
X_{ij}(w)E_{ij} )
\label{t}
\end{equation}
Performing the variation of eq.(\ref{t}) with respect to $X_{mn}(w)$ one gets
$$
\frac{\delta T_{kl}(z)}{\delta A_{mn}(w)}|_{{\cal P}_{red}}
\equiv Q^{kl}_{mn}(z,w)
=\frac1{2\pi ik}  \left( W(-w,q_{kn})
\delta_{kl}\delta_{km} -W(z-w,q_{kl})
\delta_{km}\delta_{ln}\right)
$$
By using the Poisson bracket (\ref{pb}) we get
$$
 \{g_{ij}(z),T_{kl}(w )\}|_{red}=
-L_{in}(z)Q_{jn}^{kl}(w,z).
$$
Substituting $\{ g,T\}$  and $\{ T,g\}$  brackets into 
(\ref{gh}) we can rewrite the $\{ {\cal L},{\cal L}\}$ bracket in the 
following form:  
\begin{eqnarray} 
\{{\cal L}_1(z),{\cal L}_2(w)\}|_{red} 
&=&-L_1(z)L_2(w)k^+(z,w)-k^-(z,w)L_1(z)L_2(w)\nonumber\\ &+&
L_1(z)s^-(z,w)L_2(w)+L_2(w)s^+(z,w)L_1(z),
\label{nn}
\end{eqnarray} 
where
\begin{eqnarray}
k^-(z,w)&=&-\{T_1(z),T_2(w)\},\nonumber \\
k^+(z,w)&=&\omega (z,w)-P\omega (w,z)P-\{T_1(z),T_2(w)\},\nonumber \\
s^-(z,w)&=&\omega (z,w)-\{T_1(z),T_2(w)\},\nonumber \\
s^+(z,w)&=&-P\omega (w,z)P-\{T_1(z),T_2(w)\},\nonumber 
\end{eqnarray}
and  $\omega_{ij\, kl}(z,w)=Q_{ji}^{kl}(w,z)$

It is easy to check that $Pk^{\pm}(z,w)P=
-  \delta(z-w)-k^{\pm}(w,z)$ and
$Ps^{\pm}(z,w)P=\pm s^{\mp}(w,z)$.
We also have one more important identity
$$
k^{+}(z,w)+k^{-}(z,w)=
s^{+}(z,w)+s^{-}(z,w).
$$

To complete the calculation we need the bracket 
$\{T_{ij}(z ) , T_{kl}(w )\}$ on the reduced space.
The straightforward manipulation leads to a divergent result.
By this reason we define this bracket as follows:
$$
\{T_{ij}(z ) , T_{kl}(w )\}
=\frac 1 2 \lim_{\varepsilon \to 0} \left(
\{T_{ij}(z ) , T^{\varepsilon}_{kl}(w )\}
+\{T^{\varepsilon}_{ij}(z ) , T_{kl}(w)\}\right), 
$$ 
where $T^{\varepsilon}_{kl}(z)$ is defined as a solution of 
the factorization problem with the boundary condition $T(\varepsilon 
)e=e$.  

\noindent  A simple calculation gives the following result for the bracket 
$\{T,T\}$ 
\begin{eqnarray*} 
2\pi ik\{T_{ij}(z ) , T_{kl}(w )\}  
=&
\left((\zeta (q_{ik}) - \frac{\zeta (\pi)}{\pi} q_{ik} )
(1-\delta_{ik})
+\left( \zeta (z-w) +\zeta (w )-\zeta (z )\right)
\delta_{ik}\right) \delta_{ij}  \delta_{kl} \\
+&\left( W( z-w ,q_{ik}) 
\delta_{il}\delta_{jk} + W(w,q_{ki})\delta_{il}\delta_{ij}
-W(z,q_{ik})\delta_{jk} \delta_{kl}  \right) (1-\delta_{ik})
\end{eqnarray*} 
By using this formula
we get the following expression for the coefficients:
\begin{eqnarray*} 
2\pi ik\, k^- _{ij~kl}(z,w )&=&
- \zeta (q_{ik}) \delta_{ij}\delta_{kl} (1-\delta_{ik})
- \left( \zeta (z-w) +\zeta (w )-\zeta (z )\right)
\delta_{ij}  \delta_{ik} \delta_{il}
\\
&-& \left( W(z-w ,q_{ik}) 
\delta_{il}\delta_{jk} + W(w,q_{ki})\delta_{il}\delta_{ij}
-W(z,q_{ik})\delta_{jk} \delta_{kl}  \right) (1-\delta_{ik})\\
&+& 
  \frac{\zeta (\pi)}{\pi} 
q_{ik}  \delta_{ij}\delta_{kl}
 \\
2\pi ik\, k^+ _{ij~kl}(z,y )&=&
 \left( \zeta (z-w) -\frac{\zeta(\pi)}{\pi}(z-w)   
+i\frac{z-w-\bar{z}+\bar{w}}{\Delta-\overline{\Delta}}
\right) \delta_{ij}  \delta_{ik} \delta_{il} \\
&+&\left( 
W(z-w ,q_{ik})\delta_{jk}\delta_{il} 
-\zeta(q_{ik})  \delta_{ij}\delta_{kl}\right) 
(1-\delta_{ik}) \\
&+&\frac{\zeta(\pi)}{\pi}q_{ik} 
\delta_{ij}\delta_{kl} \\
2\pi ik\, s^-_{ij~kl}(z,w)&=&
- \left( \zeta (w ) -\frac{\zeta(\pi)}{\pi}(w )   
+i\frac{w-\bar{w}}{\Delta-\overline{\Delta}}
\right) \delta_{ij}\delta_{ik} \delta_{il} \\
&-&\left( 
W(w ,q_{ki})\delta_{ij}\delta_{il} 
+\zeta(q_{ik})  \delta_{ij}\delta_{kl}\right) 
(1-\delta_{ik}) \\
&+& \frac{\zeta(\pi)}{\pi}q_{ik} 
\delta_{ij}\delta_{kl} \\
2\pi ik\, s^+_{ij~kl}(z,w)&=&
 \left( \zeta (z ) -\frac{\zeta(\pi)}{\pi}(z )   
+i\frac{z-\bar{z}}{\Delta-\overline{\Delta}}
\right) \delta_{ij}\delta_{ik} \delta_{il} \\
&+&\left( 
W(z ,q_{ik})\delta_{jk}\delta_{kl} 
-\zeta(q_{ik})  \delta_{ij}\delta_{kl}\right) 
(1-\delta_{ik}) \\
&+& \frac{\zeta(\pi)}{\pi}q_{ik} 
\delta_{ij}\delta_{kl}  \\
\end{eqnarray*}
One can easily verify
that the last lines in the expressions obtained for $k$-s and $s$-s 
do not contribute to the bracket $\{ {\cal L},{\cal L}\}$.

To proceed further, let us note that (see eq.(\ref{l})):
\begin{equation}
L_{ii} (z) = \frac 1{\sqrt{N}} W(z,h)b_i ,
\nonumber 
\end{equation}
so the bracket $\{ b_i ,b_j \}$ follows from the
$\{ L_{ii}, L_{jj}\}$ bracket only. Just as in the case of the
Heisenberg double one can  
check that the bracket of ${\cal 
L}_{ii}$ with the constraint (\ref{fe}) vanishes on ${\cal P}_{red}$ 
for any value of $\varepsilon$. Thus, there is no contribution 
from the Dirac term to the $\{ L_{ii}, L_{jj}\}$ bracket. 

Performing the same calculations as in \cite{afm} we arrive at 
\begin{equation} 
2\pi ik \{ b_i , b_j \}= b_i b_j 
(2\zeta (q_{ij})-\zeta (q_{ij}+h) -\zeta (q_{ij}-h)).  
\nonumber 
\end{equation}

The bracket $\{ {\cal L}, D\}$ and $\{ D, D\}$
can be found by a similar device as was used above.
The Dirac terms do not contribute as well.
The final result reads
\begin{equation}
\{D[A]_1,D[A]_2\}|_{red}=0,
\nonumber
\end{equation}
\begin{equation}
2\pi (\Delta-\overline{\Delta}) \{{\cal L}(z)_1,D[A]_2\}|_{red}=- 
\sum_{i,j}L_{ij}(z)E_{ij}\otimes E_{jj}. 
\label{LD} 
\end{equation}

Now for the reader's convenience we list the Poisson brackets obtained in 
terms of the coordinates on ${\cal P}_{red}$
\begin{eqnarray}
\{q_i ,q_j \}&=&0 \nonumber \\
2\pi ik  \{q_i ,b_j \}&=&b_j \delta_{ij} \nonumber \\
2\pi ik  \{ b_i , b_j \}&=& b_i b_j 
(2\zeta (q_{ij})-\zeta (q_{ij}+h) -\zeta (q_{ij}-h)),
\label{pbb}
\end{eqnarray}
this is just the Poisson structure of
the elliptic Ruijsenaars-Schneider model.

Just as in \cite{afm} the bracket for the operator $L^{ Ruij}$ 
defined by eq.(\ref{rg}) being calculated by using (\ref{nn}) and 
(\ref{LD}) reproduces the bracket obtained in \cite{Sur1}. It means that 
there is no contribution from the Dirac term even for the nondiagonal 
matrix elements $L_{ij}$.

\section{Conclusion}
In this paper we have pointed out 
that the elliptic Ruijsenaars-Schneider model can be 
obtained by means of the Hamiltonian reduction procedure
from the cotangent bundle over the two-dimensional current group.
As compare to the scheme proposed in our previous paper \cite{afm} 
this one possesses a number of advantages. First of all in this scheme 
the calculations are drastically simplified. Then, it explains why the 
contribution from the trigonometric $r$-matrix which
defines the Poisson structure on the Heisenberg double drops out from the 
final result. 

It seems to be interesting to examine the Poissonian reduction  of 
the Heisenberg double of the two-dimensional current group. In this 
case one could expect to obtain some generalization of the RS model. 

It is known that the elliptic Calogero-Moser model is related to the 
Chern-Simons theory. Hence it is an interesting problem  
to find a field-theoretical formulation of the elliptic RS model.

{\bf ACKNOWLEDGMENT} The authors are grateful to N.A.Slavnov 
for valuable discussions. This work is supported in part by the RFFR
grants N96-01-00608 and N96-01-00551 and by the ISF grant a96-1516.

\end{document}